\documentclass{camera}
\usepackage{graphicx}
\usepackage{multicol}
\usepackage{cite}
\nonumber

\begin{document}

\title{Reaction plane dispersion at intermediate energies}

\author{J. \L{}ukasik$^{1,2}$ \and W. Trautmann$^{1}$}

\organization{$^{1}$GSI, Planckstr. 1, D-64291 Darmstadt, Germany\\
$^{2}$IFJ-PAN, Pl-31342 Krak{\'o}w, Poland}

\maketitle

\begin{abstract}

A method to derive the corrections for the dispersion of the reaction plane at
intermediate energies is proposed. The method is based on the correlated,
non-isotropic Gaussian approximation. It allowed to construct the excitation
function of genuine flow values for the Au+Au reactions at 40-150 MeV/nucleon
measured with the INDRA detector at GSI.

\end{abstract}

\section{Introduction} 

Flow measurements are being performed since two decades \cite{reisdorf97,
herrmann99} starting from energies as low as few tens of MeV/nucleon up to the
ultra relativistic energies available now, and for varying system sizes and
asymmetries from C+C to Bi+Bi. Both, sideward collective motion (directed flow
related observables) and azimuthal anisotropies (elliptic flow related
observables) have been extensively studied. The primary motivation to study
these observables is their presumed link to the density and pressure attained
during the heavy ion collisions. A precise knowledge of the behavior of the
excitation function of both flow components is necessary to constrain the
transport models in order to study the properties of nuclear matter far from
the normal conditions. 

To cover a possibly large range of incident energies, different experiments
using different detectors have to be merged.  The results may depend on the
acceptance and resolution of a given apparatus as well as on the method used to
determine the flow observables and in particular the reaction plane. Thus
merging different experiments requires representing their results in a manner
least influenced by the instrumental and methodological factors. The
instrumental factors are minimized by using high resolution and high coverage
devices or correcting for inefficiencies. The method related ones, in case of
the flow observables, are minimized by presenting the values corrected for the
dispersion of the estimated reaction plane.

Since the detectors can not measure the positions, spins and momenta of the
reaction products simultaneously, the impact vector, and in particular the
reaction plane, can only be estimated using the momenta, and the precision of
the flow measurement will depend on the accuracy of this estimate. Since the
beginning of the flow investigation numerous methods have been proposed to
determine the reaction plane dispersion  \cite{danielewicz85, danielewicz88,
ollitrault92, ollitrault93, danielewicz95, ollitrault95, voloshin96,
ollitrault97, ollitrault98, poskanzer98, voloshin99, borghini00, borghini01,
borghini01a, borghini02},  however all of them are based on the central limit
theorem and thus limited to collisions at high energies which fulfill the high
multiplicity requirement. 

The established methods proved their usefulness for correcting measured flow
values at high energies (see e.g. \cite{danielewicz85, barrette94, aggarwal97,
andronic01, andronic05}). They are, however, not adequate for the intermediate
energy reactions, especially below about 100 AMeV. The present study was
motivated to fill this gap.

\section{The method}

Originally, the directed flow has been quantified by measuring the in-plane
component of the transverse momentum \cite{danielewicz85} and the elliptic flow
by parametrizing the azimuthal asymmetries in terms of the ``squeeze-out''
ratio \cite{gutbrod90}. More recently, it has been proposed to express 
flow components in terms of the Fourier coefficients, $v_{1}$, $v_{2}$ and
higher, obtained from the Fourier decomposition \cite{voloshin96, ollitrault97,
poskanzer98, borghini02} of the azimuthal distributions of the reaction
products measured with respect to the reaction plane, with azimuthal angle
$\phi_R$, and as a function of the particle type, rapidity $y$ and possibly the
transverse momentum $p_{\rm T}$:
\begin{equation}
\frac{dN}{d(\phi-\phi_R)} \propto 1+2 \sum_{n\geq1} v_n \cos n(\phi-\phi_R)
\label{eq_defvn}
\end{equation}

The first two coefficients, $v_{1} \equiv \langle\cos(\phi-\phi_R)\rangle$ and 
$v_{2} \equiv \langle\cos2(\phi-\phi_R)\rangle$, characterize the directed and
elliptic flow, respectively.

Since the azimuth of the reaction plane, $\phi_E$, can only be estimated with a
finite precision, the measured coefficients $v_{n}^{meas}$ are biased. They are
related to the genuine ones through the following expression
\cite{ollitrault97}:
\begin{equation}
v_n^{meas} \equiv \langle\cos n(\phi-\phi_E)\rangle = 
\langle\cos n(\phi-\phi_R+\phi_R-\phi_E)\rangle = 
v_n \langle\cos n\Delta\phi\rangle
\label{eq_defcorr}
\end{equation}

where the average cosine of the azimuthal angle between the true and the
estimated planes, $\langle\cos n\Delta\phi\rangle \equiv \langle\cos
n(\phi_R-\phi_E)\rangle$, is the required correction for a given harmonic.

The reaction plane can be determined by several methods, including the
flow-tensor method \cite{gyulassy82}, the fission-fragment plane
\cite{tsang84a}, the flow Q-vector method \cite{danielewicz85}, the transverse
momentum tensor \cite{ollitrault93} (``azimuthal correlation'' \cite{wilson92})
method or others \cite{fai87}. 

Taking advantage of the additivity of the Q-vector, which is defined as a weighted
sum of the transverse momenta of the measured $N$ reaction products:
\begin{equation}
\vec{Q} = \sum_{i=1}^{N}\omega_i \vec{p}_i^{\perp}
\label{eq_defqvect}
\end{equation}

we adopt this method for estimating the reaction plane. The correction factor
is then searched for using the sub-event method \cite{danielewicz85,
ollitrault97}, which consists in splitting randomly each event into two equal
multiplicity sub-events and getting the correction factor from the distribution
of the relative azimuthal angle between the Q-vectors for sub-events
(``sub-Q-vectors'') by fitting the theoretical distribution to it.

For high multiplicity events measured at high energies this theoretical
distribution can be obtained from the central limit theorem and assuming that
the sub-Q-vectors are normally distributed. Ref. \cite{ollitrault97} gives an
analytical formula for such a distribution for the case that the distributions
of sub-Q-vectors are independent and isotropic around their mean values.

At low energies the particle multiplicities are lower and the events are
characterized by a broad range of masses of the reaction products. Thus, the
applicability of the central limit theorem for devising the corrections is less
obvious. Nevertheless, taking advantage of the presumed relatively good memory
of the entrance channel plane by the heavy remnants, it can be argued that the
assumption of the Gaussian distribution of the flow Q-vector may still hold, at
least in the first approximation. QMD-CHIMERA calculations, using a version of
the code with a strict angular momentum conservation, confirm this assumption
\cite{method}, and, even more, show that the random sub-events are also
normally distributed even at 40 AMeV, except for very peripheral collisions.
The last observation is crucial since the proposed method is based on the
assumption of the normal distributions of Q-vectors for sub-events. The
difference, as compared to the high energy case, is that the sub-events are
strongly correlated \cite{ollitrault95} and that the distributions of the
Q-vector are no longer isotropic but rather elongated in the reaction plane, as
suggested in \cite{ollitrault97}. The elongation presumably reflects the
increasing role of the in-plane emissions at low energies. 

Taking into account these observations, and following the method outlined in
appendix A of \cite{borghini02}, we have derived the form of the joint
probability distribution of the random sub-Q-vectors by imposing the constraint
of momentum conservation on the $N$-particle transverse momentum distribution
and by using the saddle-point approximation.

The resulting distribution is a product of two bivariate Gaussians:
\begin{equation} 
\frac{d^{4}N}{d\vec{Q}_1d\vec{Q}_2} = 
\frac{e^{-\frac{(Q_{1x}-\bar{Q}_s)^2+(Q_{2x}-\bar{Q}_s)^2 -2\rho
(Q_{1x}-\bar{Q}_s)(Q_{2x}-\bar{Q}_s)}{\sigma_{sx}^2 (1-\rho^2)}
-\frac{Q_{1y}^2+Q_{2y}^2-2\rho Q_{1y}Q_{2y}} {\sigma_{sy}^2 (1-\rho^2)}}}{\pi^2
\sigma_{sx}^2 \sigma_{sy}^2 (1-\rho^2)}  
\label{eq_q1q2distr} 
\end{equation}

\noindent where we followed the convention of \cite{ollitrault97} of including
the $\sqrt 2$ in $\sigma$; the subscripts {\em 1, 2} and {\em s} refer to
the {\em ``sub-event''}. This distribution differs from those proposed in 
\cite{ollitrault97, ollitrault95, borghini02} in that it combines all three
effects that influence the reaction plane dispersion at intermediate energies,
namely the directed flow (through the mean in-plane component $\bar{Q}_s$ or
the resolution parameter $\chi_{s} \equiv \bar{Q}_s/\sigma_{sx}$
\cite{ollitrault97}), the elliptic flow (through the ratio $\alpha \equiv
\sigma_{sx}/\sigma_{sy}$) and the correlation between the sub-events
\cite{ollitrault95} (through the correlation coefficient $\rho \in[-1,1]$). 

Making the division into sub-events random ensures that the distributions of
the sub-Q-vectors are equivalent, in particular they have the same mean values
and variances. Since the total-Q-vector is the sum of the sub-Q-vectors,
$\vec{Q}=\vec{Q}_1+\vec{Q}_2$, its distribution takes the following form (cf
eq. (15) of ref. \cite{ollitrault97}):
\begin{equation}
\frac{d^{2}N}{dQ_xdQ_y} = 
\frac{1}{2 \pi (1+\rho) \sigma_{sx} \sigma_{sy}}
e^{-\frac{(Q_{x}-2 \bar{Q}_s)^2}{2 (1+\rho)\sigma_{sx}^2}
-\frac{Q_{y}^2}{2 (1+\rho)\sigma_{sy}^2}} 
\label{eq_qtotdistr}
\end{equation}

\noindent From this, taking into account the definition of $\chi$
\cite{ollitrault97} one finds that:
\begin{equation}
\chi = \chi_{s} \sqrt{2/(1+\rho)}
\label{eq_chi}
\end{equation}

Note that, in principle, the sub events do not have to be of equal
multiplicity, however, keeping this condition reduces the fluctuations and
makes the Gaussian assumption better fulfilled. 

As in \cite{ollitrault97} the use of the joint probability distribution
(\ref{eq_q1q2distr}) is done by integrating it over the magnitudes of the
Q-vectors and one angle, leaving the relative angle between the sub-events.
Unlike in \cite{ollitrault97}, the resulting distribution can not be presented
in an analytical form, instead, it can be calculated numerically
(two-dimensional integral) using e.g. the combined
Gauss-Legendre/Gauss-Chebyshev quadrature. It depends on 3 parameters
($\chi_{s}, \alpha, \rho$) which can be obtained from fitting it to the
experimental or model data.

The correction factors for the n-th harmonic $v_{n}$, depending now on $\chi$
and $\alpha$, can be calculated (also numerically) as the mean values of the
$\cos n\Delta\phi$ obtained over the total-Q-vector distribution
(\ref{eq_qtotdistr}), in a similar way as in \cite{ollitrault97}. 
    \vspace{-5mm}

\begin{figure}[!htb]
 \begin{minipage}[t]{0.6\linewidth}
 \leavevmode
 \begin{center}
  \includegraphics[width=0.9\linewidth]{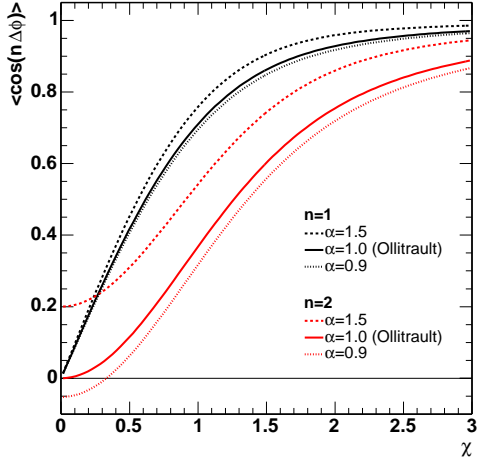}
 \end{center}
 \end{minipage}
 \hfill
 \begin{minipage}[t]{0.38\linewidth}
    \vspace*{0.9cm}
  
  \caption{Correction factors for the first (3 upper curves) and the second (3
lower curves) harmonic as a function of the resolution parameter, $\chi$, for
different aspect ratios, $\alpha$.} 

  \label{olli15}

 \end{minipage}

\end{figure}

Figure \ref{olli15} shows how the elongation of the Gaussian ($\alpha$), or
elliptic flow, modifies the correction factors for the first two harmonics. It
demonstrates that the in-plane emissions ($\alpha>1$) enhance slightly the
resolution for v$_1$ and considerably for v$_2$ -- even in the absence of the
directed flow. On the other hand, squeeze-out ($\alpha<1$) deteriorates the
resolution. In particular, it shows that the correction can be negative in case
of small directed flow and squeeze-out. The relation (\ref{eq_chi}) between
$\chi_{s}$ and $\chi$ indicates that the resolution improves in case the
sub-events are anti-correlated ($\rho<0$).

The introduction of 3 parameters in (\ref{eq_q1q2distr}) was necessary to fit
the whole family of shapes of the experimental distributions of the relative
angle between the sub-events.
\begin{figure}[!htb]
 \begin{minipage}[t]{0.6\linewidth}
 \leavevmode
 \begin{center}
  \includegraphics[width=0.9\linewidth]{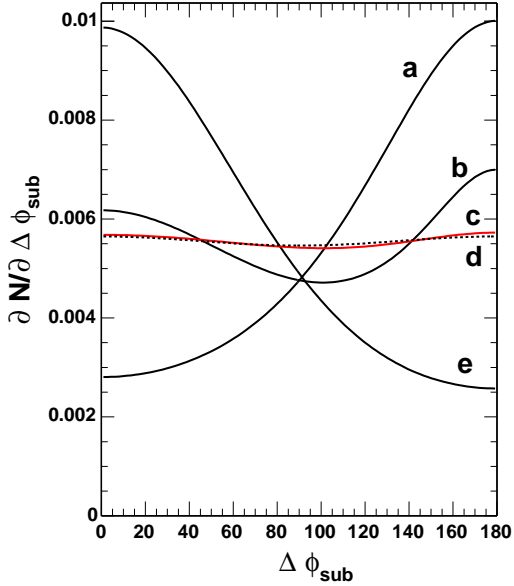}
 \end{center}
 \end{minipage}
 \hfill
 \begin{minipage}[t]{0.38\linewidth}
  
  \caption{Distribution of relative angle between the sub-Q-vectors for
different parameters together with the resulting corrections.}

\begin{tabular}{cccc}
 & $\chi$ & $\alpha$ & $\rho$ \\
a:&  0.56 & 1.00 &-0.50\\
b:&  0.91 & 1.60 &-0.40\\
c:&  0.73 & 1.20 &-0.25\\
d:&  0.00 & 1.20 & 0.00\\
e:&  1.26 & 1.20 &-0.20\\
\end{tabular} 

\vspace*{2mm}

\begin{tabular}{ccc}
 & $\langle\cos \Delta\phi\rangle$ & $\langle\cos 2\Delta\phi\rangle$\\
a:&    0.46 &  0.14\\
b:&    0.72 &  0.53\\
c:&    0.60 &  0.31\\
d:&    0.00 &  0.09\\
e:&    0.83 &  0.58\\
\end{tabular} 

  \label{shapes}

 \end{minipage}

\end{figure}

Fig. \ref{shapes} shows an example of shapes that can be described by the
integrated distribution (\ref{eq_q1q2distr}). The parameters used for each
curve are specified in the caption together with the corresponding correction
factors. Case (a) corresponds to events with small directed flow and strong
anti-correlation, producing a pronounced backward peaking observed at low
energies and more central collisions. Case (b) corresponds to less central
collisions at low energies. Case (e) corresponds to mid-central collisions at
intermediate energies, where the directed flow increases. At higher energies
the peak at small relative angles becomes narrower. 

Examples (c) and (d) deserve a special comment. They are almost
indistinguishable. Case (d) represents an almost isotropic distribution with no
directed flow and for independent sub-events ($\chi = 0$, $\rho=0$) with the
vanishing correction for directed flow ($\langle\cos \Delta\phi\rangle=0$),
while (c) represents the case with non-zero directed flow and moderate
anti-correlation ($\rho=-0.25$). In this case the corrections are substantially
different from zero. This example shows the difficulties that can be
encountered while fitting the experimental distributions. In these special
cases the method may lose its resolving power, unless one can find a way to
constrain some of the parameters, e.g. $\rho$. One way to do this is to get a
hint on the value of the correlation coefficient from model calculations. For
example, the CHIMERA calculations predict this coefficient to be around -0.43
for 40 AMeV and $2<b<8$ fm and about -0.2 at 150 AMeV, indicating
that the case (d) with vanishing flow, is less likely to represent the
experimental distribution. Alternatively, one can impose constraints by
calculating some rotational invariants from the measured data, e.g. $\langle
\vec{Q}_1 \cdot \vec{Q}_2 \rangle$ and $\langle Q_1^2 \rangle$ or $\langle
Q_2^2 \rangle$. From these the following ratio can be constructed: 

\begin{equation}
\frac{\langle \vec{Q}_1 \cdot \vec{Q}_2 \rangle}{\langle Q_1^2 \rangle}  
\equiv \beta = 
\frac{\rho+\alpha^2 (\rho + 2 \chi_{s}^2)}{1+\alpha^2 (1 + 2\chi_{s}^2)} 
\label{eq_beta}
\end{equation}

This relation allows to reduce the number of fit parameters to two by using the
experimentally accessible ratio $\beta$. 

For the cases (c) and (d) from the previous example $\beta$ takes the values of
-0.0114 and 0, respectively. Taking into account that $\beta$ can be known with
high accuracy for high statistics data (a few percent at worst), its precise
knowledge constrains the fitting routine to search for a conditional minimum
and helps to resolve the ambiguous cases.

Instead of fitting the relative angle distributions one can express the
probability distribution (\ref{eq_q1q2distr}) in terms of projections of one
sub-Q-vector in the reference frame of the other. The corresponding
2-dimensional experimental distributions can then be fit using such a formula.
The drawback might be that this formula depends on the magnitudes of the
sub-Q-vectors, which introduces a dependence not only on the experimental
uncertainties of the measured angles but also on the accuracy of the energy
calibration. The advantage is that it can be calculated using only
one-dimensional numerical integration. This formula contains four parameters,
with $\sigma_{sx}$ being the additional one. However, using the calculated
invariants, it is again possible to reduce the number of fit parameters to
two.

\section{Preliminary results}

The experimental data unavoidably suffers from inaccuracies and lacking
completeness. The fitting procedure yields relatively accurate results for the
corrections for the first two harmonics in case of the simulations (e.g. 2-5\%
for 40 AMeV and 0.2-0.4\% for 150 AMeV and $4<b<8$ fm), but in the case of the
experimental data it will certainly return some effective correction factors
biased by the experimental uncertainties.

In order to minimize the experimental uncertainties one can either restrict the
data sample to ``quasi-complete'' events, with the possibility that it is no
longer representative, or one can try to complete the event information by
applying conservation laws, with the possibility of generating additional
fluctuations.

In the following, we adopt the latter approach. We analyze all the events which
contain at least 35\% of the total charge and complete them by adding a single
``extra'' fragment carrying the missing momentum, charge and mass. This
procedure changes substantially the distributions of the relative angle between
sub-events for peripheral collisions where, due to the energy thresholds, the
heavy target-like fragment is always lost. The distributions become narrowly
peaked at small relative angles which improves the resolution of the reaction
plane.

\begin{figure}[!htb]
 \leavevmode
 \begin{center}
  \includegraphics[width=0.9\linewidth]{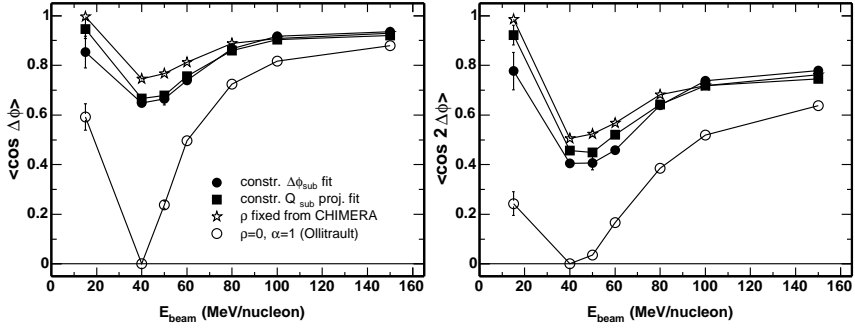}
 \end{center}
  
  \caption{Correction parameters for $v_1$ (left) and $v_2$ (right) for $4\leq
  b \leq 6$ fm collisions of Au+Au measured with the INDRA detector.} 

  \label{corrpar}
\end{figure}

Fig. \ref{corrpar} presents the correction parameters for the first two
harmonics calculated on the basis of the parameters $\chi_{s}, \alpha, \rho$
obtained from the fits to the experimental distributions of the relative angle
between sub-events (solid circles), and from the fits of the projections of one
sub-Q-vector on the other (solid squares). In both cases the calculated
rotational invariants were used to constrain the fits. The stars represent the
results of the fits with the fixed $\rho$ as obtained from the CHIMERA
calculation, and the open circles represent the results obtained using the
standard Ollitrault method \cite{ollitrault97}, which corresponds to the case
of $\rho=0$ and $\alpha=1$. The figure demonstrates the utility and accuracy of
the proposed method in cases where the standard methods do not apply, i.e. at
low energies. It shows that the standard and the extended method approach each
other at higher energies, but even at 150 AMeV there is still about 5\% and
15\% discrepancy for the $v_1$ and  $v_2$ corrections, respectively. Since at
150 AMeV the elliptic flow is small, this discrepancy comes mostly from
correlations between the sub-events which are still non-negligible ($\rho
\simeq -0.2$) at this energy. Differences between the fit results obtained
using the angular distributions and the projections, based on the new method
(solid symbols), reflect its systematic uncertainty which, apart from the 15
AMeV case, can be quantified as about 2\% for directed flow and up to about
10\% for the elliptic flow.

\begin{figure}[!htb]
 \leavevmode
 \begin{center}
  \includegraphics[width=0.9\linewidth]{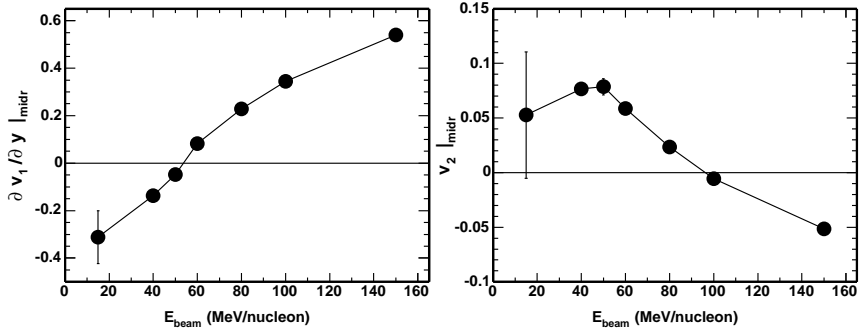}
 \end{center}

  \caption{Corrected slopes of $v_1$ at midrapidity (left) and $v_2$  at
  midrapidity (right) for Z=2 particles from mid-central ($4\leq b \leq 6$ fm)
  collisions of Au+Au measured with the INDRA detector.} 

  \label{v1v2}
\end{figure}

Figure \ref{v1v2} shows the excitation function of the slope of $v_1$ at
midrapidity (left) and of $v_2$ at midrapidity (right) corrected for the
dispersion of the reconstructed reaction plane using the corrections obtained
with the present method (angular distribution fits).

The $v_1$ and $v_2$ parameters have been calculated in the beam frame,
excluding the particle of interest from the reaction plane reconstruction and
correcting for the momentum conservation \cite{borghini02}. The linear fits to
obtain the slope parameters have been carried out in the range of $\pm 0.4$ of
the scaled CM rapidity for all energies except 15 AMeV, where the range of $\pm
0.55$ has been used to improve the accuracy. In the case of the second harmonic
the value of $v_2$ at midrapidity has been obtained by fitting a parabola to
the $v_2$ vs scaled rapidity relation and taking the value of the fit at
$y_{CM}=0$. The point at 15 AMeV has been included in the systematics to get a
hint on the evolution of the excitation function at very low energies. Due to
the very low statistics, this result should be taken with care and, possibly,
verified with dedicated and more complete measurements. 

The trends observed for the uncorrected data \cite{lukasik05} for $v_1$ are
preserved. The excitation function for Z=2 changes sign between 50 and 60 AMeV
and continues to decrease at lower energies.

The excitation function of the elliptic flow shows a possible saturation of the
in-plane enhancement near 0.08 for bombarding energies below about 60 AMeV or
even a signature of a maximum near 50 AMeV.

\section{Summary} 

We have proposed a new method to calculate the corrections resulting from the
dispersion of the reaction plane at intermediate energies. The method, which in
some respect is an extension of the one proposed by Ollitrault
\cite{ollitrault97}, takes into account the combined effects of the directed
flow, elliptic flow and correlations on the reaction plane resolution. It shows
an increasing role of correlations and of the in-plane enhancement at lower
energies. The corrected experimental excitation functions of $v_1$ and $v_2$ 
constitute a precise constraint for the dynamical models which are aimed at
inferring the properties of nuclear matter at varying densities and pressures
from flow observables.

\end{document}